\documentclass[12pt,a4paper]{article}
\pdfoutput=1
\usepackage{jcappub}
\usepackage[english]{babel}
\usepackage{multirow}
\usepackage{ulem}
\usepackage{url}
\usepackage[utf8]{inputenc}

\def\HI{\hbox{H\,{\sc i}}}
\def\bea{\begin{eqnarray}}
\def\eea{\end{eqnarray}}
\def\be{\begin{equation}}
\def\ee{\end{equation}}
\newcommand{\red}{\color{black}}

\newcommand{\Eq}[1]{Eq.~\eqref{#1}}
\newcommand{\Fig}[1]{Fig.~\ref{#1}}

\newcommand{\Sec}[1]{Sec.~\ref{#1}}

\title{The EMU view of the Large Magellanic Cloud: Troubles for sub-TeV WIMPs}

\author[a,b]{Marco Regis,}
\emailAdd{marco.regis@unito.it}
\author[a,b]{Javier Reynoso-Cordova,}
\author[c]{Miroslav D. Filipovi\'c,}
\author[d]{Marcus Br\"{u}ggen,}
\author[e]{Ettore Carretti,}
\author[f,c]{Jordan Collier,}
\author[g,c]{Andrew M. Hopkins,}
\author[h]{Emil Lenc,}
\author[i]{Umberto Maio,}
\author[l]{Joshua R. Marvil,}
\author[c,h]{Ray P. Norris,}
\author[m]{and Tessa Vernstrom}
\affiliation[a]{Dipartimento di Fisica, Universit\`{a} di Torino, via P. Giuria 1, I--10125 Torino, Italy}
\affiliation[b]{Istituto Nazionale di Fisica Nucleare, Sezione di Torino, via P. Giuria 1, I--10125 Torino, Italy}
\affiliation[c]{Western Sydney University, Locked Bag 1797, Penrith South DC, NSW 2751, Australia}
\affiliation[d]{University of Hamburg, Gojenbergsweg 112, 21029 Hamburg, Germany}
\affiliation[e]{INAF Istituto di Radioastronomia, via Gobetti 101, 40129 Bologna, Italy}
\affiliation[f]{University of Cape Town, The Inter-University Institute for Data Intensive Astronomy (IDIA), Department of Astronomy, Private Bag X3, Rondebosch 7701, South Africa}
\affiliation[g]{Australian Astronomical Optics, Macquarie University, 105 Delhi Rd, North Ryde, NSW 2113, Australia}
\affiliation[h]{CSIRO, Space and Astronomy, PO Box 76, Epping, NSW 1710, Australia}
\affiliation[i]{INAF  - Italian National Institute for Astrophysics, Observatory of Trieste, via G. Tiepolo~ 11, 34143 Trieste, Italy}
\affiliation[l]{National Radio Astronomy Observatory, P.O. Box O, Socorro, NM 87801, USA}
\affiliation[m]{CSIRO Astronomy and Space Science, Kensington Perth 6151, Australia}

\abstract{
We present a radio search for WIMP dark matter in the Large Magellanic Cloud (LMC).
We make use of a recent deep image of the LMC obtained from observations of the Australian Square Kilometre Array Pathfinder (ASKAP), and processed as part of the Evolutionary Map of the Universe (EMU) survey.
LMC is an extremely promising target for WIMP searches at radio frequencies because of the large J-factor and the presence of a substantial magnetic field. 
We detect no evidence for emission arising from WIMP annihilations and derive stringent bounds {\red on the annihilation rate as a function of the WIMP mass, for different annihilation channels}. This work excludes the thermal cross section for masses below 480 GeV and annihilation into quarks.
}

\date{\today}
\begin{document}
\maketitle

\section{Introduction}
\label{sec:Intro}
The nature of dark matter (DM) is one of the defining mysteries of modern physics.
One of the most attractive candidates proposed to solve the DM puzzle is given by {\red hypothetical} particles that are more massive than baryons and weakly interacting, the so-called WIMPs~\cite{Bertone:2010zza}.

WIMPs have masses in the GeV-TeV range and can annihilate in pairs into lighter particles.
In particular, electrons and positrons can be directly or indirectly injected by WIMP annihilations, and a sizable final branching ratio of annihilation into $e^+-e^-$ is a rather generic feature of WIMP models (see, e.g., Fig. 4 in Ref.~\cite{Regis:2008ij}). 
Emitted in an environment with a background magnetic field, high-energy electrons and positrons give rise to synchrotron radiation typically peaking at radio frequencies.

The Large Magellanic Cloud (LMC) is the most massive satellite galaxy of the Milky Way. 
The large dark matter {\red mass (with the LMC virial mass being around $10^{11}\,M_\odot$~\cite{Kallivayalil_2013})} and the proximity to Earth (distance around 50 kpc~\cite{Clementini_2003}) make the LMC one of the best targets for indirect searches of WIMPs.
The so-called J-factor, namely, the integral of the density squared over the line-of-sight and solid angle, amounts to $\sim 10^{20}{\rm GeV^2/cm^5}$~\cite{Buckley:2015doa}, second only to the Galactic center.
The presence in LMC of a $\mu$G magnetic field~\cite{Gaensler:2005} suggests an investigation of a possible WIMP signature at radio frequencies.

The idea of deriving bounds on WIMPs from radio observations of the LMC is not new~\cite{Tasitsiomi:2004,Siffert:2011}. The improvement presented in the current analysis with respect to Refs.~\cite{Tasitsiomi:2004} and \cite{Siffert:2011} arises from new, more sensitive, data, an updated model (with the inclusion of spatial diffusion in the computation of the DM signal), and the choice of statistical approach (comprising a morphological analysis {\red with pixel by pixel comparison of the observed and predicted signals}).

We use observations of the LMC obtained by the Australian Square Kilometre Array Pathfinder (ASKAP \cite{2021PASA...38....9H}), as part of the ASKAP commissioning and early science (ACES, project code AS033) verification and made available to the Evolutionary Map of the Universe (EMU) project~\cite{Norris:2011}. These data were used to obtain a deep image of the LMC at 888 MHz~\cite{Pennock:2021}, which will be the starting point of the analysis of this work. 

The structure of the paper is as follows. Section~\ref{sec:obs} describes the ASKAP observations and how the LMC radio image has been created. The model of DM and interstellar medium in the LMC is presented in Section~\ref{sec:model}. We introduce the statistical analysis and present the results in Section~\ref{sec:res}, while Section~\ref{sec:comp} provides a comparison with previous works. Conclusions are drawn in Section~\ref{sec:conc}. The Appendix is devoted to consistency checks.

\section{Observational Maps}
 \label{sec:obs}

\begin{figure}[ht!]
   \includegraphics[width=0.49\textwidth]{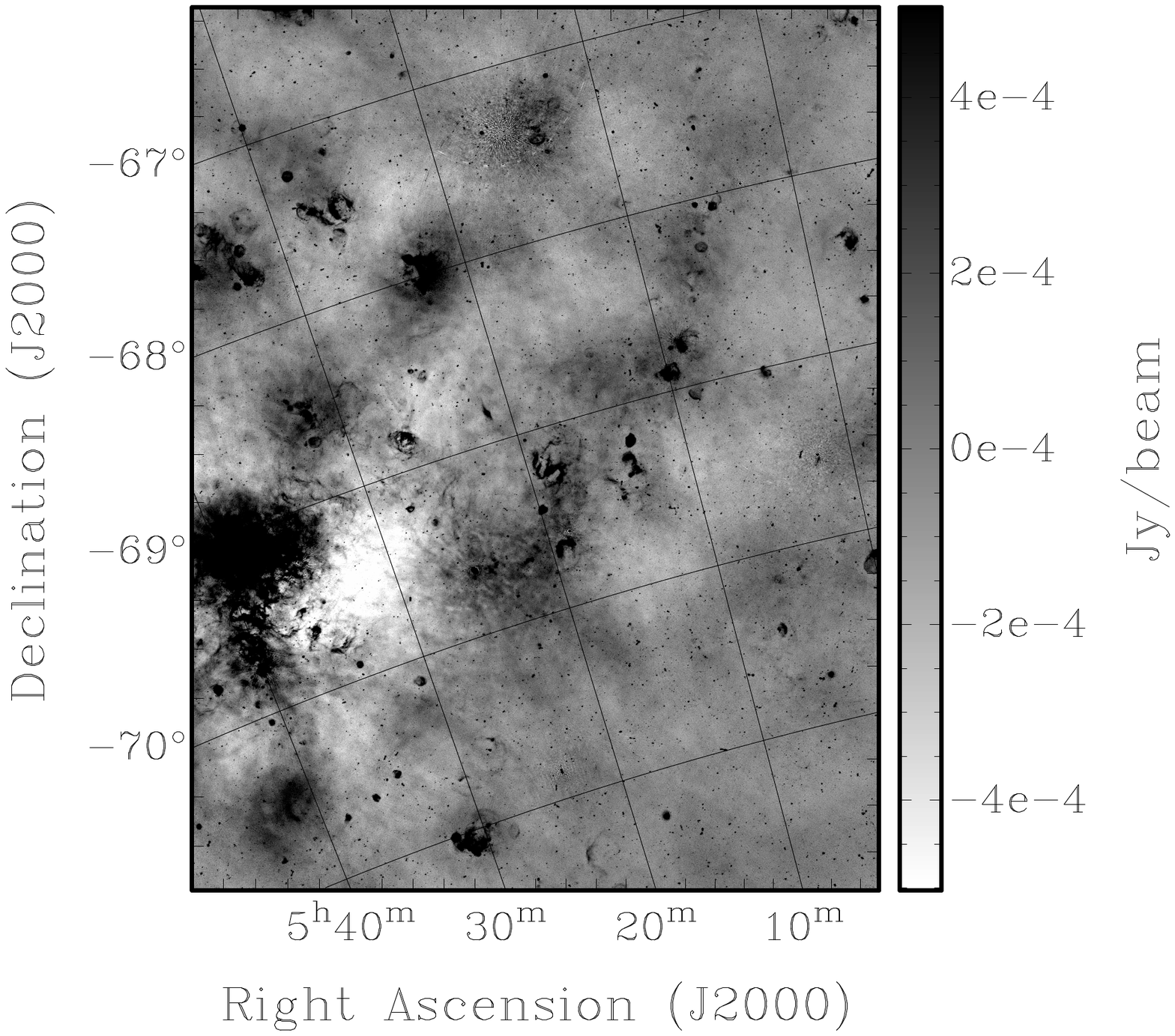}
   \includegraphics[width=0.49\textwidth]{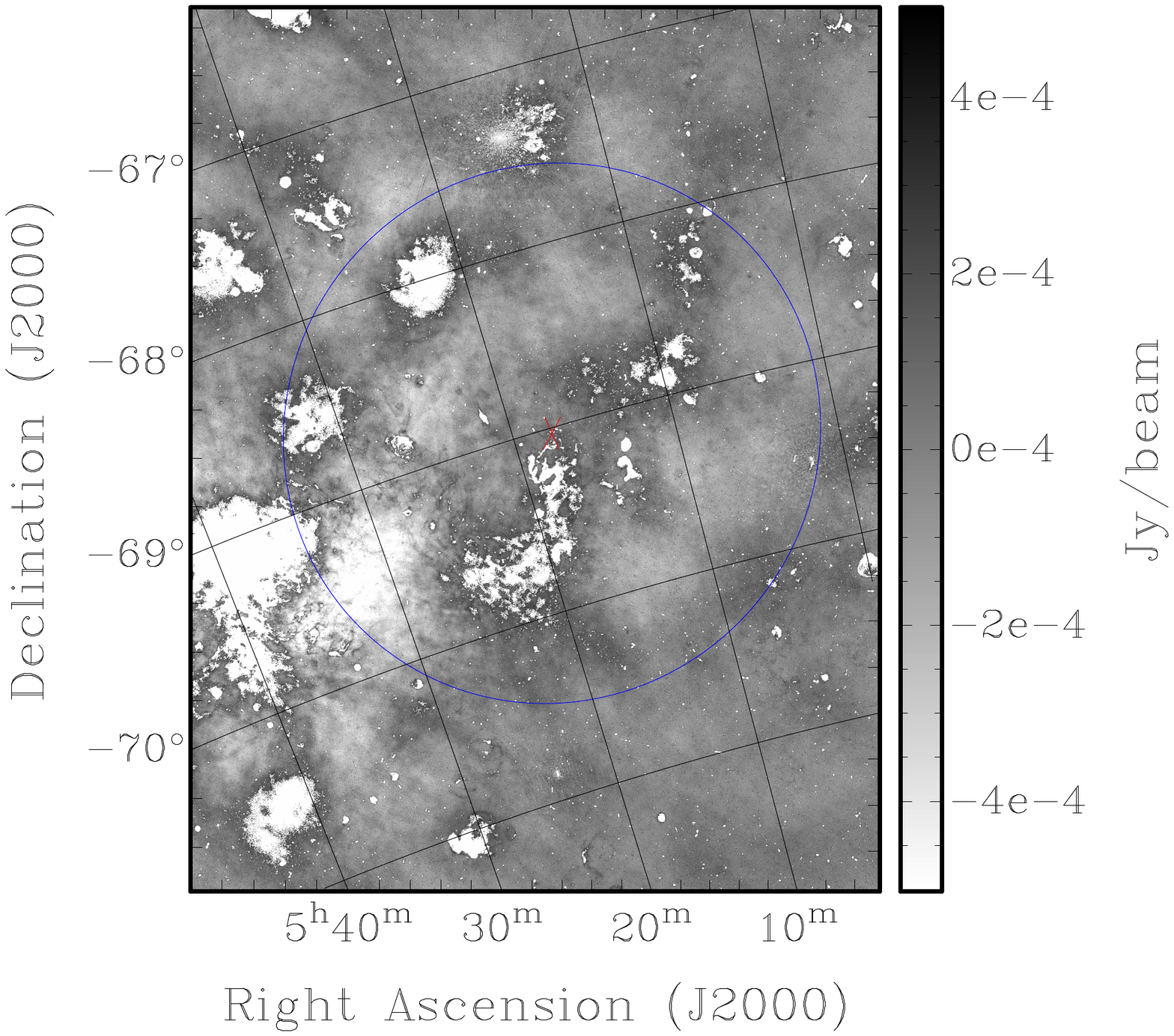}
   \vspace{-2cm}
    \caption{{\bf Left panel:} Observational map {\red of the LMC at 888 MHz obtained from ASKAP data in} Ref.~\cite{Pennock:2021}.
    {\bf Right panel:} Same as left panel but blanking masked pixels. The red cross indicates the position of the dynamical center and the blue circle encloses the region of interest for this work. }
\label{fig:maps}
 \end{figure}

In this work we make use of the observations of the LMC taken as part of the ASKAP commissioning and early science at 888 MHz with a bandwidth of 288 MHz, and analysed as part of the EMU project \cite{Pennock:2021}. The observations cover a total field of view of 120 deg$^2$, with a total exposure time of about 12h40m. Data processing was performed using ASKAPsoft by the ASKAP operations team and the resulting images are available on the CSIRO ASKAP Science Data Archive. The beam size of the map shown in \Fig{fig:maps} (left) is $13.9'' \times 12.1''$ and the median Root Mean Squared (RMS) noise is $\sim 58\,\mu$Jy/beam. For more details see Ref.~\cite{Pennock:2021}.

Structures on scales $\lesssim 2^\circ$ can be recovered thanks to the short baselines of the ASKAP array (with shortest one being 22~m). We expect the DM-induced emission to be diffuse but showing variations on scales below $2^\circ$ (see next Section).
We confirm the sensitivity of the image to DM diffuse emissions by performing a test detailed in the Appendix.

{\red Our search looks for a possible diffuse emission associated to the LMC halo, and all the small-scale discrete sources (in the LMC or in the background/foreground) are a contaminant that we attempt to mask. }
We {\red identify} discrete sources by running the publicly available tool SExtractor~\cite{Bertin:1996fj}, which is also used to derive the RMS map, in the same way as described in Ref.~\cite{Regis:2014koa}. The threshold for a source to be masked is set to $3\times RMS$, and the result is shown in \Fig{fig:maps} (right). We also {\red mask} negative pixels using the same criterion, i.e., absolute value larger than $3\times RMS$. They are likely due to missing short-spacing data. 

Since the expected emission from DM has a size of several arcmin, we further smooth the masked image (using the task SMOOTH in {\it Miriad}~\cite{Sault:95}) to FWHM$=2^\prime$, in order to be more stable against small-scale residuals and fluctuations.

\section{LMC description}
\label{sec:model}
We compute the radio emission induced by WIMP DM by combining the synchrotron power associated with the LMC magnetic field with the equilibrium distribution ($n_e$) of electrons and positrons injected by DM. In order to compute $n_e$, we solve a transport equation describing the cooling and spatial diffusion experienced by the electrons and positrons after injection.
We describe it in the limit of spherical symmetry and stationarity:
\be
 -\frac{1}{r^2}\frac{\partial}{\partial r}\left[r^2 D\frac{\partial f}{\partial r} \right] 
  +\frac{1}{p^2}\frac{\partial}{\partial p}(\dot p p^2 f)=
  s( r, p)\;,
\label{eq:transp}
\ee
where $f(r,p)$ is the $e^+/e^-$ distribution function at the equilibrium,\footnote{We assume equilibrium since the timescales associated to diffusion and cooling are around 10-30 Myr (see below), much smaller than the age of the LMC (around 1 Gyr).} at a given radius $r$ (from the LMC dynamical center) and at a given momentum $p$. The distribution $f$ is related to the number density in the energy interval $(E,E+dE)$ by: $n_e(r,E)dE=4\pi \,p^2f(r,p)dp$; analogously, for the $e^+/e^-$ source function, we have $q_e(r,E)dE=4\pi \,p^2\,s(r,p)dp$. 
The first term on the left-hand side describes the spatial diffusion, with $D(r,p)$ being the diffusion coefficient. The second term accounts for the energy loss due to radiative processes; $\dot p(r,p)=\sum_i dp_i(r,p)/dt$ is the sum of the rates of momentum loss associated to the radiative process $i$.
The DM source $q_e$ scales with the number density of WIMP pairs locally in space, i.e., with $\rho^2/(2\,M_{\chi}^2)$, where $\rho(r)$ is the halo mass density profile, and $M_{\chi}$ is the mass of the DM particle.\footnote{In the case of WIMP as a Dirac fermion, {\red the number density of WIMP pairs goes as $\rho^2/(4\,M_{\chi}^2)$, while $\rho^2/(2\,M_{\chi}^2)$ } is appropriate for the more common cases of WIMP as a boson or Majorana fermion.} We neglect substructure contributions and assume the DM spatial distribution to be spherically symmetric and static.
The source term associated to the production of $e^+/e^-$ is given by:
\be
q^{a}_e(E,r)=\langle\sigma_a v\rangle\,\frac{\rho(r)^2}{2\,M_{\chi}^2} \times \frac{dN_e^{a}}{dE}(E) \;,
\label{eq:QDM}
\ee
where $\langle \sigma_a v\rangle$ is the velocity-averaged annihilation rate, and $dN_e^a/dE$ is the number of electrons/positrons emitted per annihilation in the energy interval $(E,E+dE)$ for a given annihilation channel.

We solve \Eq{eq:transp} numerically using finite-differencing Crank-Nicolson scheme, for details see Ref.~\cite{Regis:2014koa}. Boundary conditions are set to be Neumann's one at the centre and Dirichlet's one at the farthest boundary, the latter chosen to be ten times the radius of our region of interest (RoI).
A recent semi-analytical treatment of \Eq{eq:transp} can be found in Ref.~\cite{Vollmann:2020}.

The properties we want to constrain are the DM mass and annihilation rate, while the ingredients we need to model are the DM spatial profile, the magnetic field, the spatial diffusion coefficient, the CMB and LMC interstellar radiation fields (ISRF, for inverse Compton losses) and the gas density (for bremsstrahlung losses), that we describe in detail in the following sections.
Since our goal is to derive conservative bounds on the WIMP signal, we will model the above quantities taking lower limits for the DM profile and magnetic field, while upper limits for diffusion coefficient and ISRF and gas densities.

To limit uncertainties in the model description, our RoI for the analysis will be defined by a relatively small region around the LMC center, corresponding to 1.3 kpc in radius ($1.5^\circ$ in angular units).
The loss in terms of J-factor, if compared to considering the full LMC halo, is limited, around a factor of two, depending on the profile. 
{\red As we will describe in the following, in such RoI we have a more robust determination of the various ingredients entering the computation, such as the magnetic field, the gas and ISRF distributions, and the DM profile. Moreover, we exclude the bulk of the contamination from the 30 Doradus region (south-west in \Fig{fig:maps}).}

\subsection{Dark matter profile}
\label{sec:DMprof}

\begin{figure}[ht!]
\centering
\includegraphics[width=0.8\textwidth]{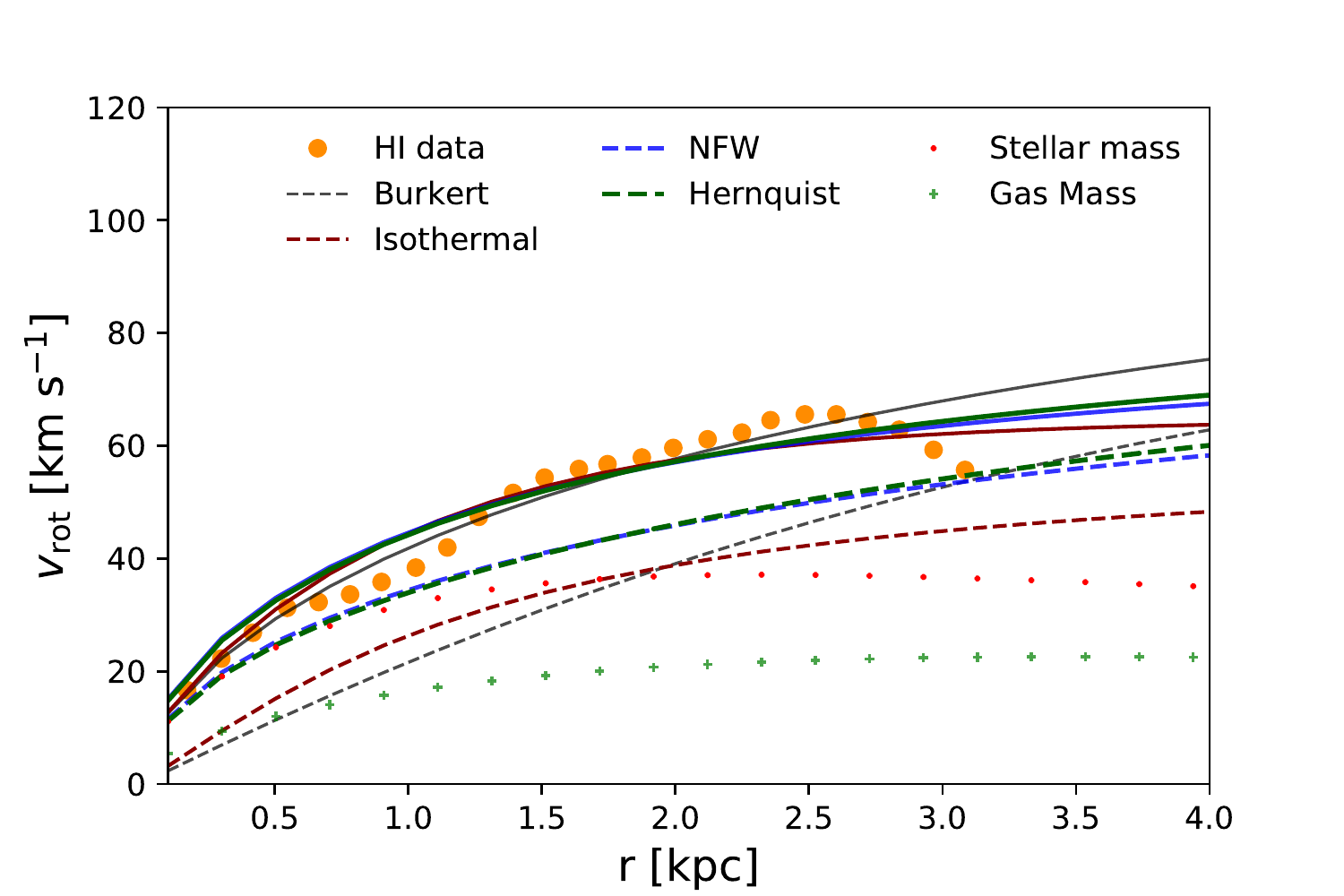}
\caption{LMC rotation curve. The \HI\  rotation data points are shown as orange dots \cite{Kim:1998}. Dashed lines show the contribution from the enclosed DM mass for the different profiles assumed (see Section \ref{sec:DMprof}). Red dots and green crosses denote the stellar and gas contributions respectively. Solid lines correspond to the contribution of all the enclosed mass within a radius $r$ (DM+stellar+gas).}
\label{fig:v_rot}
\end{figure}

To model the radio emission of the LMC due to annihilating DM, the spatial particle distribution $\rho(r)$ is a key ingredient, see \Eq{eq:QDM}. 
Previous work has explored different functional forms for the DM density profile in the LMC \cite{PhysRevD.91.102001,Buckley:2015doa,Siffert:2011,PhysRevD.93.062004,Besla_2019,TASITSIOMI2004637,Garavito_Camargo_2019} and analysed \HI\ rotational \cite{Kim:1998} and carbon star data \cite{van_der_Marel_2014}  to constrain the parameters of $\rho(r)$. In this work we are interested in the inner region, so we make use of the \HI\ data \cite{Kim:1998} that provide the most accurate rotation velocities at small distances from the LMC dynamical center. We explore four different profiles and fit $\rho(r)$ up to a radius of $\sim 2.7$ kpc from the center\footnote{{\red We discard the last points in \Fig{fig:v_rot}, since they might be affected by systematic errors, mainly due to non-circular motions \cite{Kim:1998} .}}, which corresponds to about twice our RoI.
In particular, we consider two different ``cuspy'' DM profiles from the NFW model (Navarro-Frenk-White) \cite{Navarro:1995iw} and Hernquist \cite{1990ApJ...356..359H}:
\be
\rho_{\rm{NFW}}(r)=\frac{\rho_s}{\left(\frac{r}{r_S}\right)\left( 1 + \frac{r}{r_s} \right)^2}\;\;\;,\;\;\;\rho_{\rm{Her}}(r)= \frac{\rho_s}{\left( \frac{r}{r_s}\right)\left(1+\frac{r}{r_s}\right)^3}\;,
\ee
and two ``cored'' profiles, the isothermal sphere \cite{1991MNRAS.249..523B} and the Burkert profile \cite{Burkert_1995}:
\be
\rho_{\rm{Iso}}(r)=\frac{\rho_s}{ 1 + \left( \frac{r}{r_s}\right)^2}\;\;\;,\;\;\; \rho_{\rm{Bur}}(r)=\frac{\rho_s}{\left( 1 + \frac{r}{r_s}\right)\left(1+ \left(\frac{r}{r_s}\right)^2\right)}\;.
\ee
The main reason for considering different shapes is related to our poor knowledge about DM physics at small scales, including the possible role of baryons in affecting the DM spatial profile. The range of possibilities encompassed by the above functional forms should bracket the uncertainty.

The free parameters of the different profiles are the scale radius $r_s$ and the normalization $\rho_s$. We fit these values for each profile using the \HI\  rotation velocity data and the fact that the velocity $v(r)$ measured up to a radius $r$ is given by the expression 
\begin{equation}
    v(r) = \sqrt{\frac{G M_{\rm{tot}}(r)}{r}},
\end{equation}
where $M_{\rm{tot}}(r)$ is the total mass contained within a radius $r$, given by DM plus contributions from the stellar and gas components. We model the stellar potential $\phi_{\star}(R,z)$ using a Plummer-Kuzmin disk \cite{1975PASJ...27..533M} {\red as a function of the disk radial distance $R$ and vertical height $z$ in cylindrical coordinates:} 
\begin{equation}
    \phi_{\star}(R,z)= GM_{\star}\left[R^2 + \left(a_{\star} + \sqrt{z^2 + b_{\star}^2}\right)^2 \right]^{-1/2},
\end{equation}
where $a_{\star}$ and $b_{\star}$ are the radial scale length and vertical scale height, respectively, for which we take $a_{\star}=1.7$ kpc and $b_{\star}=0.34$ kpc \cite{Salem2015,Bustard:2020}. The stellar mass $M_{\star}$ is left as a free parameter in the fit. The contribution to the mass density $\rho_g(R,z)$ from the gas follows the expression given in  Ref. \cite{Bustard:2020} (with radial scale length $a_g=a_{\star}$ and vertical scale height $b_g=b_{\star}$):
\begin{equation}
\rho_g(r,z)= \frac{M_g}{2\pi a_g^2 b_g}\, 0.5^2\, {\rm sech}\left(\frac{R}{a_g}\right){\rm sech}\left(\frac{|z|}{b_g}\right).
\label{eq:gas}
\end{equation}
Once we have the total mass contribution from the different components to the rotation velocity at some radius $r$, we proceed to fit the parameters $r_s$, $\rho_s$ and $M_{\star}$ using a least squares method through the python package \texttt{scipy.optimize}.  The best-fit parameters for the different DM density profiles are shown on Table \ref{Table:DM_parameters}.
{\red For simplicity, and since the gas component provides a subdominant contribution to the matter density, we set $M_g=5\times 10^8\,M_\odot$. Nevertheless, we checked that the results for $r_s$ and $\rho_s$ reported on Table \ref{Table:DM_parameters} are unchanged if $M_g$ is left as a free parameter in the fit.}

Previous work has suggested that the LMC virial mass is around $2\times10^{11} M_\odot$, but estimates can have significant uncertainties  \cite{Kallivayalil_2013,Garavito_Camargo_2019,Besla_2019}. On the other hand, the {\red virial mass is mostly related to the DM profile at larger radii than the one relevant for our analysis (the enclosed mass in our RoI is $\lesssim 2\times10^{10} M_\odot$), making these uncertainties of little relevance for our results.} We adopt different values for the LMC virial mass, corresponding to Table 2 from Ref.~\cite{Garavito_Camargo_2019}, in order to normalize the profiles. We find that the rotation velocity within $\sim 5$ kpc from the center of the LMC, as well as the profile parameters determined by our fit, do not change considerably with different choices for the mass normalization. Therefore we adopt a (low) virial mass of $10^{11} M_\odot$.

\begin{table}[h]
\centering
 \begin{tabular}{||c c c c ||} 
 \hline
 Profile & $r_s$ [kpc] & $\rho_s$ $[M_\odot/\rm{kcp}^3 ]$& $M_{\star}$ $[M_\odot ]$  \\ [0.5ex] 
 \hline\hline
 NFW & 9.8 & $5.1\times10^{6}$ & $1.0\times10^{9}$ \\ 
 \hline
 Isothermal & 1.1 & $5.7\times10^{7}$& $1.9\times10^{9}$  \\
 \hline
 Burket & 4.7 & $3.1\times10^7$& $1.9\times10^{9}$  \\
 \hline
 Hernquist & 21.8 & $2.1\times10^{6}$& $1.0\times10^{9}$ \\ [1ex] 
 \hline
\end{tabular}
\caption{Parameters for the LMC DM density profiles {\red and stellar mass derived from the rotation curve fit}, see Section \ref{sec:DMprof}.}
\label{Table:DM_parameters}
\end{table}

The results are shown in Figure \ref{fig:v_rot}, where we report the contributions to the rotation velocity from DM (dashed lines), stellar and gas components (red dots and green crosses respectively). The orange points represent the \HI\  rotation data from Ref.~\cite{Kim:1998} and the solid lines show the contribution from the sum of all components.

\subsection{Magnetic field and diffusion coefficient}
\label{sec:B}
The strength of the large-scale coherent component of the LMC magnetic field is found to be $1\,\mu$G, as determined via Faraday rotation measure of polarized background sources~\cite{Gaensler:2005}, and with diffuse polarized data~\cite{Mao:2012} (see also \cite{FaradaySky:2021}).
The turbulent component is expected to be larger than the regular one, by a factor $> 3$, as generically found in galaxies (see, e.g., Ref.~\cite{Beck:2019}), and confirmed by the scatter observed in the LMC rotation measures~\cite{Mao:2012}. In Ref.~\cite{Gaensler:2005}, the total LMC magnetic field strength on large scales has been estimated to be $B=4.3\,\mu$G, and we take this as our reference value. We focus on a relatively small and central region of LMC, where rotation measures do not show significant dependence on the radial distance, so we can assume a uniform strength, in agreement with the model in Ref.~\cite{Mao:2012}.  Amplifications on small scales~\cite{Gaensler:2005} are disregarded.
A recent analysis based on the equipartition assumption and on observations of the LMC synchrotron emission at 166 MHz~\cite{Hassani:2021} suggests a higher value, $B=7.7\pm 1.1\,\mu$G.
Since the magnetic field is (together with the DM properties) the most crucial ingredient of our analysis, we show how our results change for a range of total strength $B=2-8\,\mu$G. This range is in agreement with estimates stemming from the cosmic-ray density derived from $\gamma$-ray data and again applying the equipartition assumption~\cite{Mao:2012}.

Data on supernovae remnants~\cite{Bozzetto:2017} and on large-scale diffuse emission~\cite{Murphy:2012} indicate that the transport of cosmic-rays in the LMC proceeds in a similar way as in other nearby galaxies, and can be explained as diffusive propagation in a turbulent regime. In this scenario, the diffusion coefficient can be estimated as~\cite{Murphy:2012}:
\be
D\simeq 3\,\times 10^{27}\,\left( \frac{d_L}{{\rm kpc}}\right)^2\,\left( \frac{10^{15}\,{\rm s}}{\tau}\right) {\rm \frac{cm^2}{s}}\;,
\label{eq:D0}
\ee
where $d_L$ is the diffusion length and $\tau$ is the cooling time.
Radio observations suggest $d_L\simeq 1-2$ kpc in the vertical direction (larger along the disk), consistent with findings in other galaxies where the confinement region is a few times the disk height. The discussion on radiative losses below leads to $\tau\simeq 10^{15}$ s, in agreement with the limit of $\tau >10^{14}$ s estimated in Ref.~\cite{Bozzetto:2017}.

Thus observations point towards a value somewhat lower than in the Galaxy.
For clarity, and in the spirit of making conservative assumptions, we assume the same strength and energy dependence of the diffusion coefficient as at large scales in the Galaxy, taking the latest determination from Ref.~\cite{Weinrich:2020cmw} (BIG model, which provides $2\,\times 10^{28}\,{\rm cm^2/s}$ at 4 GeV). 
Recall that the larger the diffusion coefficient the smaller the DM signal, since diffusion can remove electrons and positrons from the RoI before they emit synchrotron radiation at the frequency of interest.

\subsection{Gas and interstellar radiation fields}
\label{sec:losses}
We determine the central value of the gas density from \Eq{eq:gas} and taking $M_g=5\times 10^8\,M_\odot$, which is the neutral hydrogen mass observed by Ref.~\cite{Kim:1998}. Then we assume a flat spatial profile, normalized to the maximal value, i.e., the central value, which leads to {\red the gas number density} $n_g=\rho(0,0)/m_H=0.8\,{\rm cm^{-3}}$ {\red (where $m_H=0.938$ GeV is the hydrogen atom mass)}. This is clearly a conservative approach. Moreover it simplifies the computation by avoiding uncertainties related to the radial and vertical scale lengths of the gas distribution and allowing us to keep assuming spherical symmetry. 
We checked this description by deriving the gas density from the hydrogen column density in Fig.4 of Ref.~\cite{Kim:2003} divided by a disk thickness of 350 pc~\cite{Kim:2003}. The spatial profile does not show significant variations in our RoI (justifying a flat model) and the average value for the column density is $6\times 10^{20}\,{\rm cm}^{-2}$, which translates into $n_g\simeq 0.5\,{\rm cm^{-3}}$, confirming the above assumption for $n_g$ as an upper limit.

We assume the gas to be composed solely of neutral atomic hydrogen, since molecular hydrogen and ionised gas are subdominant components~\cite{1999PASJ...51..745F,2002ApJ...566..857T,Gaensler:2005}, negligible in this analysis.

The ISRF spectrum is taken to have the same shape as that of the Milky-Way~\cite{Porter:2008ve}. Observationally, this is found to be a good approximation~\cite{Bernard:2008}. Moreover, even though we implement a full computation, the Klein-Nishina corrections are subdominant (the energy of the emitting electrons is $\lesssim 10$ GeV), so the exact ISRF spectral shape is not critical, and the size of the inverse Compton losses is essentially set by the integral over energy.

The normalization is chosen such that the integral of the spectrum provides $U_{\rm{ISRF}}=1\,{\rm eV/cm^3}$, consistent with the parameter $X_{\rm{ISRF}}$ found in Ref.~\cite{Bernard:2008}. We note that a somewhat lower density can be derived from the LMC SED of Ref.~\cite{Israel:2010}, see Ref.~\cite{Foreman:2015} who found $U_{\rm{ISRF}}= 0.57\,{\rm eV/cm^3}$, and using the estimate of Ref.~\cite{Murphy:2012} where $U_{\rm{ISRF}}\simeq 0.3\,{\rm eV/cm^3}$. Again, our choice is in the spirit of adopting a realistic upper limit.

As for the gas density, we conservatively take a spatially flat profile.

\begin{figure}[ht!]
\vspace{-2.cm}
\begin{minipage}{10cm} 
   \includegraphics[width=0.8\textwidth]{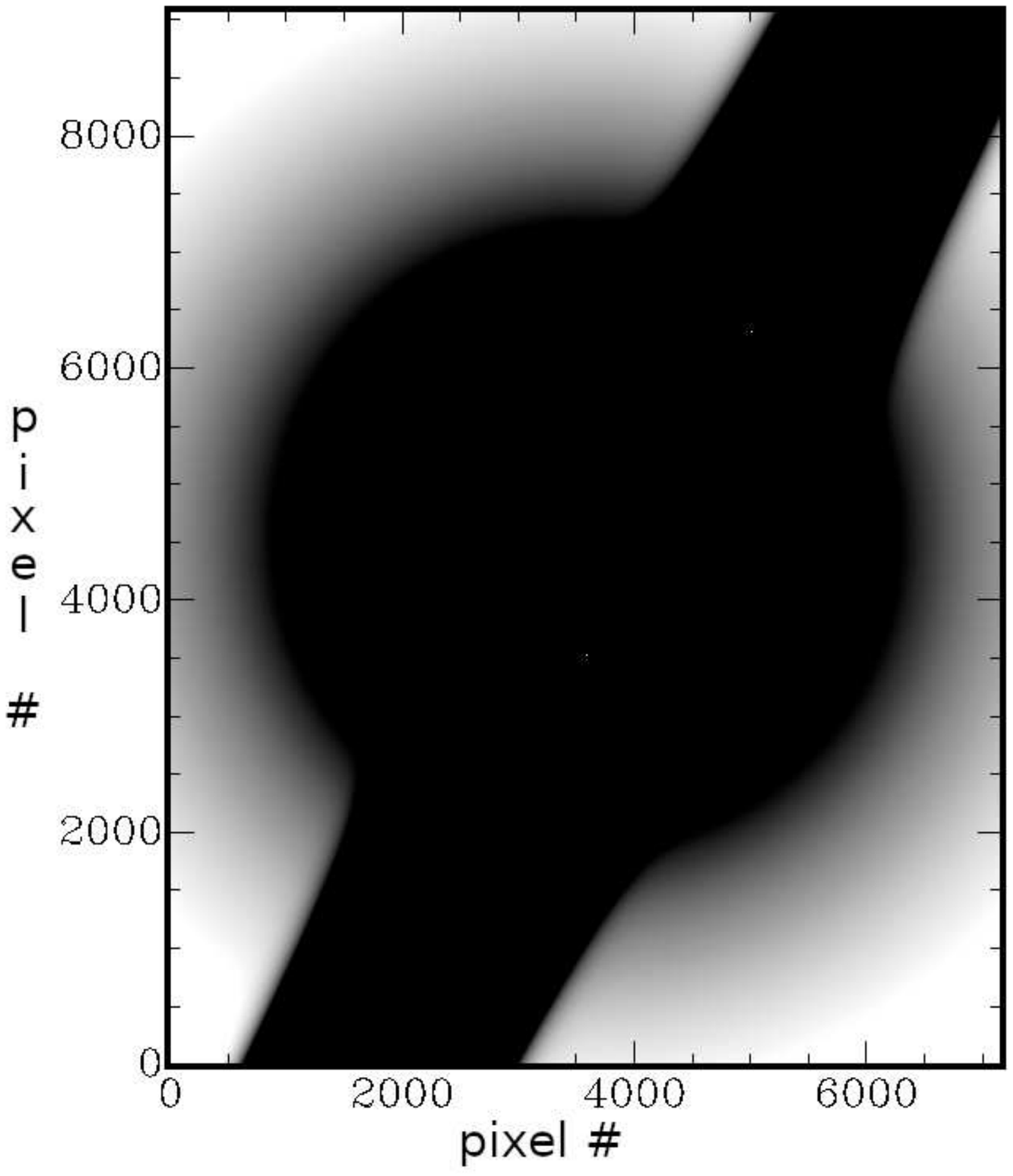}
 \end{minipage}
 \hspace{-2cm}
 \begin{minipage}[htb]{8cm}
   \centering
   \includegraphics[width=\textwidth]{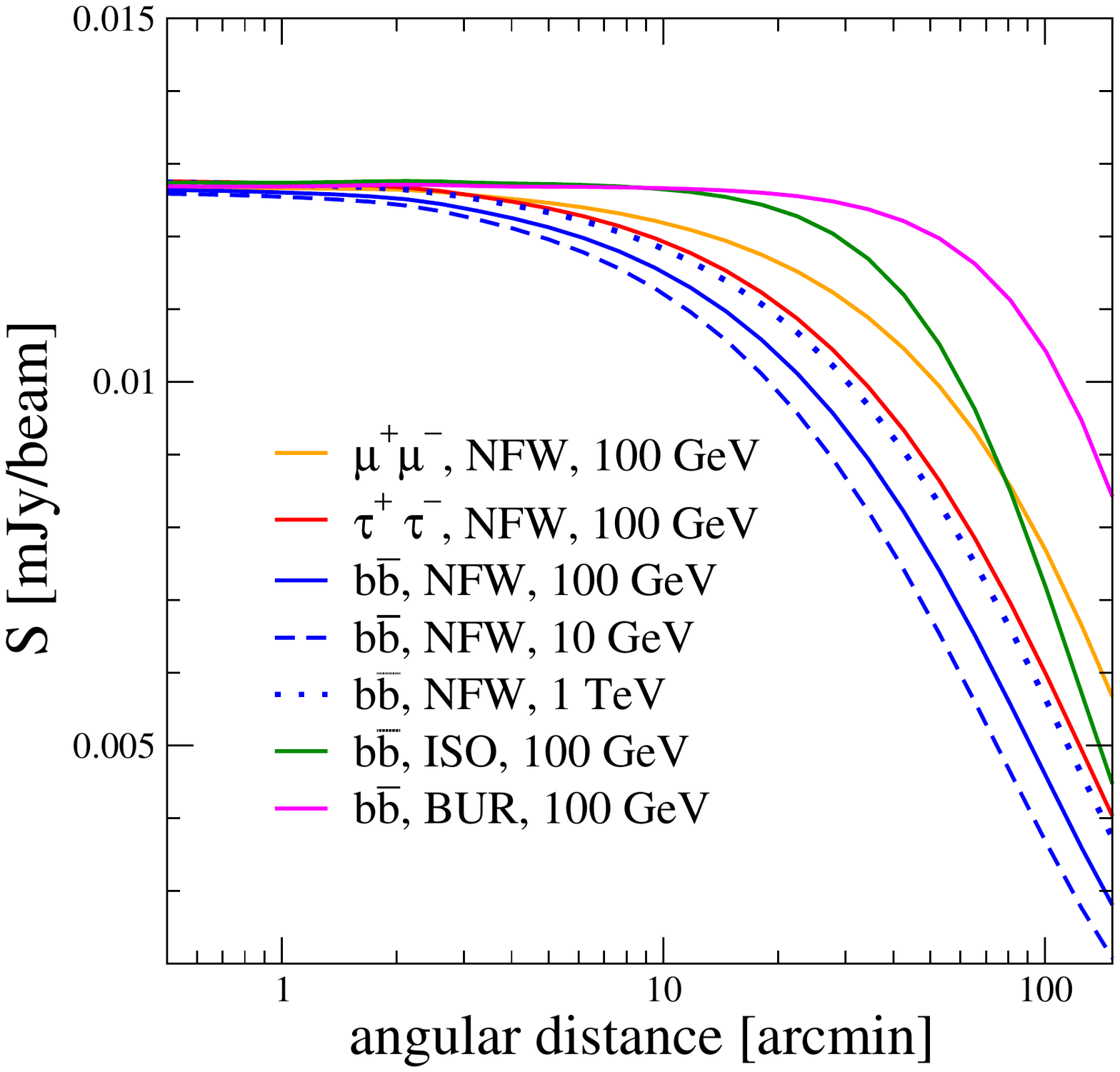}
 \end{minipage}
    \caption{{\bf Left panel:} Model of LMC diffuse emission: disk+DM.
    {\bf Right panel:} WIMP emission profile as a function of the angular distance from the LMC center, for different WIMP masses, annihilation channels, and spatial profiles {\red (with the magnetic field strength taken to be $B=4.3\,\mu$G)}.
}
\label{fig:emprof}
 \end{figure}

\section{Results}
\label{sec:res}

We assume the likelihood associated with the LMC diffuse emission to be described by a Gaussian:
\be 
\mathcal{L}=e^{-\chi^2/2} \;\;\; {\rm with} \;\;\; \chi^2=\frac{1}{N_{pix}^{FWHM}}\sum_{i=1}^{N_{pix}} \left(\frac{S_{th}^i-S_{obs}^i}{\sigma_{rms}^i}\right)^2\;,
\label{eq:like}
\ee
where $S_{th}^i$ is the theoretical estimate for the flux density in the pixel $i$, $S_{obs}^i$ is the observed flux density and $\sigma_{rms}^i$ is the r.m.s. error, both described in \Sec{sec:obs}. 
$N_{pix}$ is the total number of pixels in the RoI (excluding masked pixels) and $N_{pix}^{FWHM}$ is the number of pixels within the FWHM of the synthesized beam.
We only include the DM signal coming from inside a sphere of {\red radius of} 1.3 kpc around the LMC center (thus disregarding other line-of-sight DM contributions inside the angular region of $90^\prime$).
The theoretical estimate is provided by the WIMP emission, computed from the solution of \Eq{eq:transp} and following Sec. 4.1 in Ref.~\cite{Regis:2014koa}, plus a disk component and a spatially flat term.

The disk is described through a Gaussian $S_{disc}^i=S_0\,\exp{[-\theta_{d,i}^2/(2\,\theta_0^2)]}$, where 
$\theta_{d,i}$ is the angular distance of the pixel $i$ from the axis of the disk, and $S_0$ and $\theta_0$ are free parameters.
The position angle of the LMC disk has been found to be between $122.5^\circ$ and $170.5^\circ$, depending on the tracer (see Ref.~\cite{2021AAS...23755210K} and references therein). In our analysis we assume a value determined by fitting the map without including the DM component. We find $138^\circ$, similar to that found recently using Gaia DR2 data~\cite{2019MNRAS.490.1076E}.

In \Fig{fig:emprof} (left), we show the shape of the model given by disk plus DM. We use arbitrary normalization and fix the FWHM of the Gaussian describing the disk to $0.45^\circ$ (which is the best-fit value found in the fit).

On top of the disk component we add a spatially flat term $S_{flat}$. This is included in the fit to account for possible offsets or a large-scale foreground component. The parameters $S_{flat}$, $S_0$ and $\theta_0$ are treated as nuisance parameters.

In \Fig{fig:emprof} (right), we show the WIMP emission as a function of the angular distance from the center, for different masses, annihilation channels and DM density profiles. Note that the size of the DM source is below 2 degrees.
The NFW profile is the most concentrated case (together with the Hernquist profile, which is not shown since it is nearly identical to the NFW at small distances). The Burkert and isothermal profiles provide {\red more extended} distributions. Note that high masses and leptonic channels imply a {\red less concentrated} profile than low masses and hadronic channels. This can be understood by noting that at the frequency of the observations (888 MHz) and for a magnetic field of $4.3\,\mu$G, the synchrotron emission is mostly provided by $e^+/e^-$ with energy around a few GeV.
High energy electrons take time to cool down to few GeV and thus can travel long distances, flattening the central overdensity.
Recall that leptonic channels provide harder $e^+e^-$ spectra than in the $b\bar b$ case.

Bounds on the parameter $\langle \sigma_a v \rangle$ are computed at any given mass $M_\chi$
through a profile likelihood technique~\cite{2005NIMPA.551..493R}, namely ``profiling out'' the nuisance parameters $\vec\Pi=(S_{flat},S_0,\theta_0)$.
We assume that $\lambda_c(\langle \sigma_a v\rangle)=-2\ln[\mathcal{L}(\langle \sigma_a v\rangle,\vec\Pi^{b.f.})/\mathcal{L}(\langle \sigma_a v\rangle^{b.f.},\vec\Pi^{b.f.})]$ follows a $\chi^2$-distribution with one d.o.f.\ and with one-sided probability given by $P=\int^{\infty}_{\sqrt{\lambda_c}}d\chi\,e^{-\chi^2/2}/\sqrt{2\,\pi}$, where $\langle \sigma_a v\rangle^{b.f.}$ denotes the best-fit value for the annihilation rate at that specific WIMP mass.
Therefore, the 95\% C.L. upper limit on $\langle \sigma_a v\rangle$ at mass $M_\chi$ is obtained by increasing the signal from its best-fit value until $\lambda_c=2.71$, keeping $\vec\Pi$ fixed to its best-fit value.

The possible presence of a DM signal is investigated by evaluating the difference $\Delta \chi^2=\chi^2(\vec\Pi^{b.f.},\langle \sigma_a v\rangle=0)-\chi^2(\vec\Pi^{b.f.},\langle \sigma_a v\rangle_{b.f.})$,  We always find $\Delta \chi^2<1$, and thus no evidence for a diffuse component associated with WIMP-induced emission.

\begin{figure}[ht!]
\vspace{-2.cm}
   \includegraphics[width=0.49\textwidth]{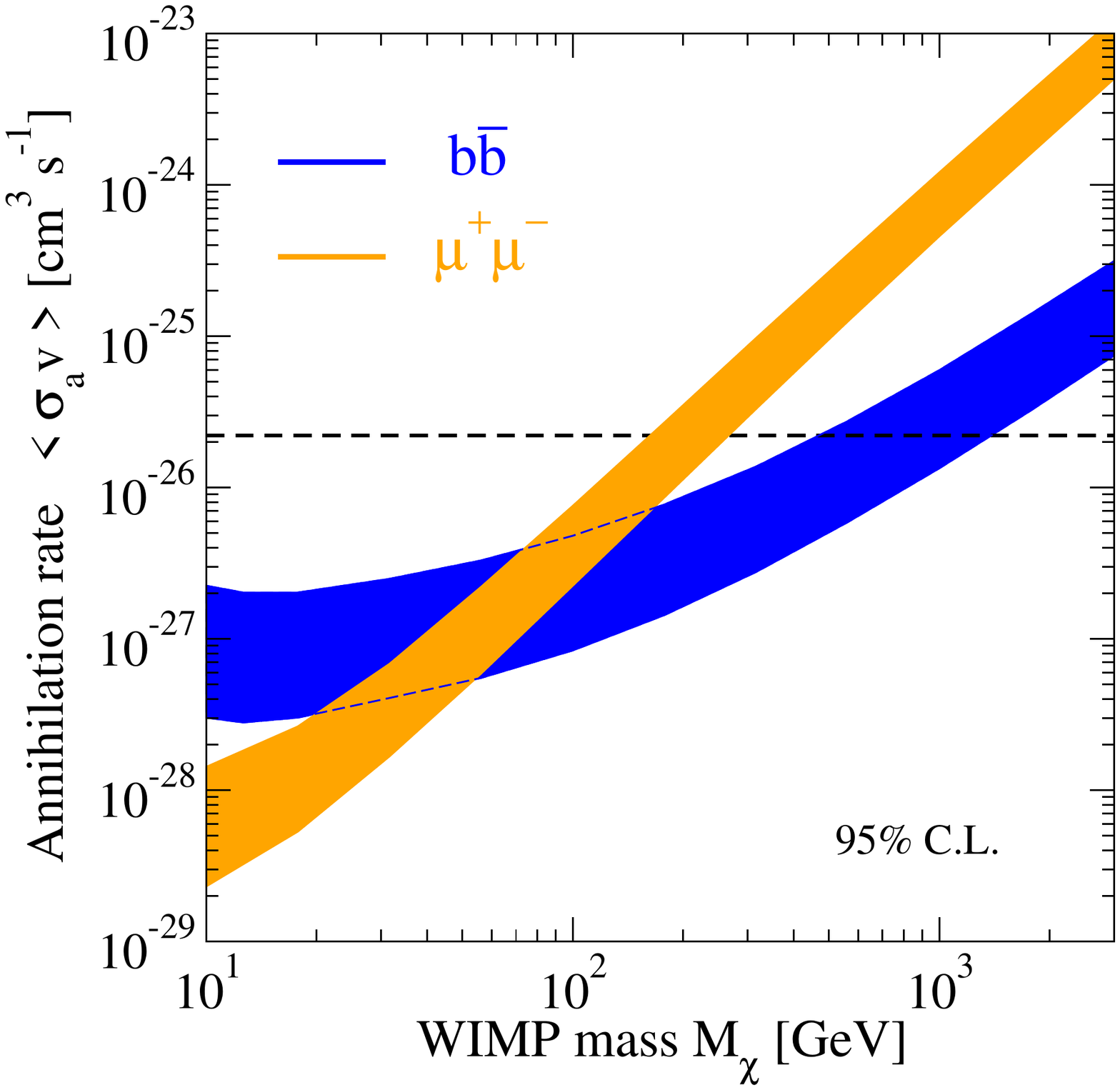}
   \includegraphics[width=0.49\textwidth]{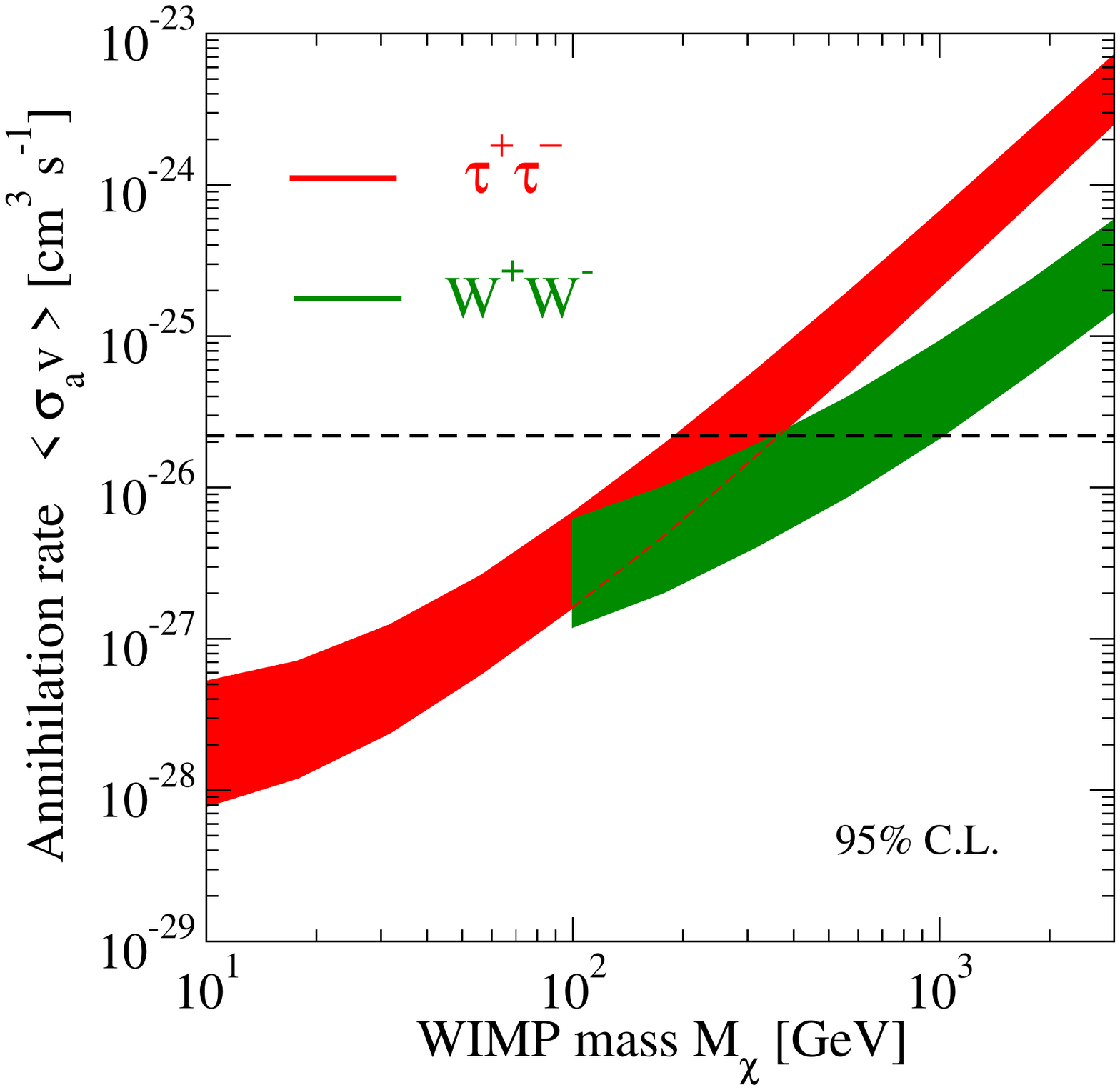}
    \caption{95\% C.L. upper limits on $\langle \sigma_a v\rangle$ as a function of $M_\chi$ for the annihilation channels $b\bar b$ (blue), $\mu^+\mu^-$ (orange), $\tau^+\tau^-$ (red), and $W^+W^-$ (green), {\red with the magnetic field strength taken to be $B=4.3\,\mu$G}.
}
\label{fig:resmain}
 \end{figure}

Results are shown in \Fig{fig:resmain}, reporting the upper limits on $\langle \sigma_a v \rangle$. The boundaries of the uncertainty band are determined by taking the weakest/strongest bound among those obtained using the four different DM profiles described in Section~\ref{sec:DMprof}. More concentrated profiles provide more stringent constraints. The NFW and Hernquist cases set the lower boundary of the band, while Burkert at low masses and Isothermal at high masses set the upper boundary.
The dashed black line is the so-called thermal cross section, namely the self-annihilation cross section needed in the early Universe in order to provide the DM mass density observed today~\cite{Steigman:2012nb}. A common way to see \Fig{fig:resmain} is to consider ``canonical'' WIMPs excluded for masses where the bound is below the thermal value.

The trend of the bound is similar for the $b\bar b$ (blue) and  $W^+W^-$ (green) channels, on one hand, and for $\tau^+\tau^-$ (red) and $\mu^+\mu^-$ (orange) channels, on the other. The reason is related to the injection spectrum of $e^+e^-$. Let us first remind that the key quantity is the density of $e^+/e^-$ induced by the specific DM model at energies of a few GeV. The injection spectrum of $e^+e^-$ is harder in the leptonic cases, where WIMPs with mass of tens of GeV have therefore a significant injection of $e^+e^-$ with energy around the peak of the synchrotron power. This makes the bounds in the $\tau^+\tau^-$ and $\mu^+\mu^-$ cases very tight at low masses. Clearly, the picture is the opposite at high masses where the injection energy is ``too high'' and the $e^+e^-$ undergo energy losses and diffusion before emitting synchrotron radiation.
In the cases of $b\bar b$ and  $W^+W^-$, the peak in the injection of $e^+e^-$ occurs at around $M_\chi/20$. Therefore, they are more efficiently constrained in the range of masses around hundreds of GeV (so that, again, the production of $e^+/e^-$ is peaked around a few GeV).

{\red Since we consider non-relativistic DM, the WIMP mass has to be larger than the mass of the annihilation products, and this is the reason of the cut in the $W^+W^-$ bound.}

The overall increase of the constraints with the WIMP mass, occurring for all the channels, can be understood from \Eq{eq:QDM}.

The bottom line of \Fig{fig:resmain} is that the thermal cross-section is excluded for masses below (480, 358, 192, 164) GeV for the ($b\bar b$, $W^+W^-$, $\tau^+\tau^-$, $\mu^+\mu^-$) annihilation channel.

\begin{figure}[ht!]
\vspace{-2.cm}
   \includegraphics[width=0.49\textwidth]{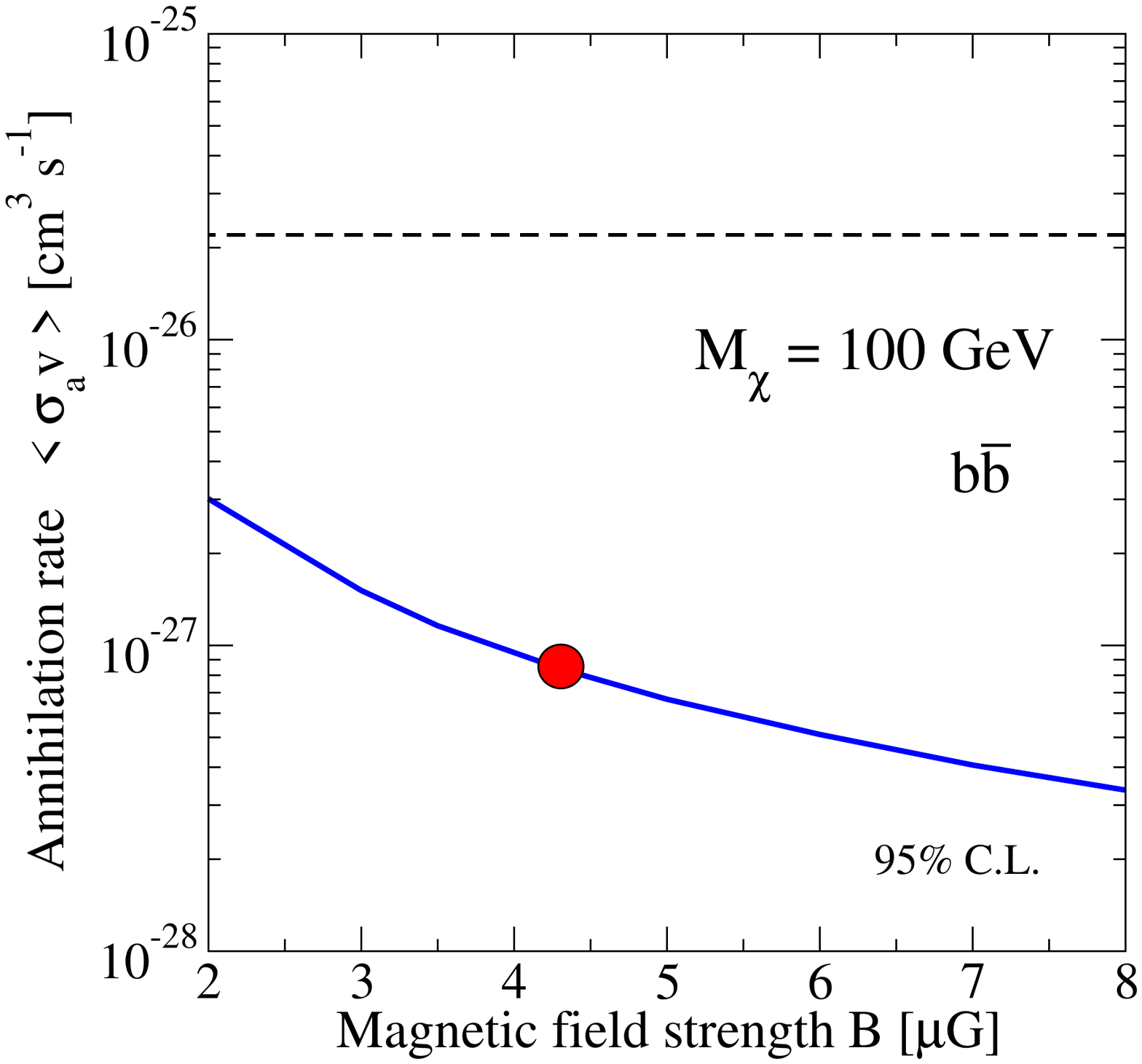}
   \includegraphics[width=0.49\textwidth]{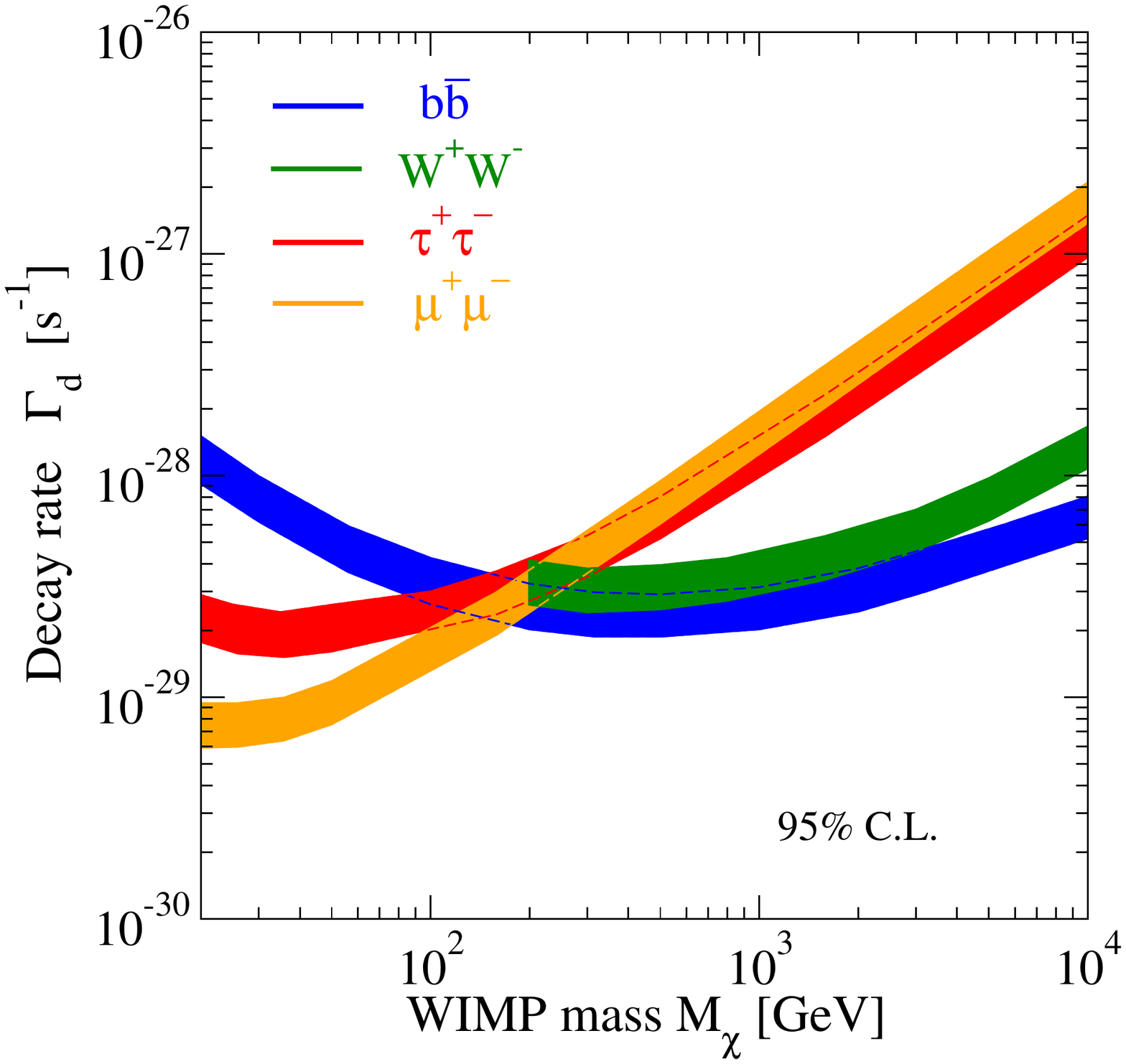}
    \caption{{\bf Left panel:}  Impact of the magnetic field strength on the bound on $\langle \sigma_a v\rangle$, in an example taking $M_\chi=100$ GeV, annihilation into $b\bar b$, and NFW DM profile. The red dot shows the reference value adopted in this work.
    {\bf Right panel:} 95\% C.L. upper limits on the decay rate $\Gamma$ as a function of $M_\chi$ for the decay channels $b\bar b$ (blue), $\mu^+\mu^-$ (orange), $\tau^+\tau^-$ (red), and $W^+W^-$ (green), {\red with the magnetic field strength taken to be $B=4.3\,\mu$G}.}
\label{fig:res2}
 \end{figure}

In \Fig{fig:res2} (left), we show how the magnetic field strength affects the bound on $\langle \sigma_a v\rangle$. We consider an example with $M_\chi=100$ GeV, annihilation into $b\bar b$, and NFW DM profile.
The bound approximately scales with the inverse of the square of the magnetic field for small strengths, and flattens as the strength increases. As already stated, in \Fig{fig:resmain} we adopted $B=4.3\,\mu$G.

Throughout the paper, we have been assuming annihilating DM. In the right panel of \Fig{fig:res2} we show the bounds that can be obtained for decaying DM. The only difference from the above analysis consists in the replacement of \Eq{eq:QDM} with 
\be
q^{d}_e(E,r)=\Gamma_d\,\frac{\rho(r)}{M_{\chi}} \times \frac{dN_e^{d}}{dE}(E) \;,
\label{eq:Qdec}
\ee
where $\Gamma_d$ is the decay rate, and $dN_e^d/dE$ is the number of electrons/positrons emitted per decay in the energy interval $(E,E+dE)$.

The different behaviour of the four decaying channels can be understood in a very similar way to that already discussed above for the annihilating case.
Note that the uncertainty band of the curves in \Fig{fig:res2} is smaller than in the annihilating cases of \Fig{fig:resmain}.
This is because the source function of annihilating DM depends on $\rho^2$, while the decaying scenario scales linearly with $\rho$, and thus uncertainties in the DM spatial profile affect the former more than the latter.

\section{Comparison with previous work}
\label{sec:comp}
We focus the comparison with previous analyses on work investigating the LMC and dwarf galaxies, i.e., satellites of the Milky Way, since they share a similar analysis as that conducted here. We do not attempt to make comparisons with completely different targets (e.g., the Galactic Center) or channels (e.g., antiprotons).
For other analyses using radio data to constrain WIMP annihilations in extragalactic objects, see, e.g., Refs. \cite{Egorov:2013exa,McDaniel:2018,ChanM31} (M31), \cite{Borriello:2009tt,Chan:2017ric} (M33), \cite{Colafrancesco:2005ji,Storm:2013} (clusters), \cite{Fornengo:2011xk,Colafrancesco:2014coa} (cosmological emission).
\subsection{Comparison with radio analyses}
An analysis similar to the one presented here was conducted in Ref.~\cite{Siffert:2011}. They employed ATCA+Parkes data and obtained the black curve in \Fig{fig:comp} (left). 
The great improvement in the constraining power of our analysis is not due to the model, for which we adopted a more conservative description than Ref.~\cite{Siffert:2011}, but to the different statistical approach and to the more limited residuals and higher rms sensitivity in the ASKAP image compared to the ATCA+Parkes image. Concerning the statistical analysis, Ref.~\cite{Siffert:2011} derived the bound from individual lines of sight, while we developed a morphological analysis. The latter allows us to ascribe part of the LMC emission to a disk component and combines $\sim 10^4$ lines of sight. This is important since the constraining power scales roughly as the square root of this number. 

There have been a few attempts to detect WIMP-induced radio signals in dwarf spheroidal galaxies of the Local Group. We expect the signal from LMC to be stronger than from dSphs since the magnetic field strength is higher (it is actually unknown in dSph, but typically assumed to be around 1 $\mu$G~\cite{Regis:2014koa}), the J-factor is higher than (or at the level of) that in the most promising dSphs, and the LMC is bigger (which means diffusion effects are less relevant in depleting the signal than in dSphs). In \Fig{fig:comp}, we include bounds {\red derived from the observations of different samples of dSphs} with the ATCA from Ref.~\cite{Regis:2017} (taking their model with $B=1\,\mu{\rm G}\;,\;D_0=3\,\times 10^{28}\,{\rm cm^2/s}$, red solid line) and Ref.~\cite{Regis:2014tga} (AVE model, red dashed line), GMRT~\cite{Basu:2021} ($B=2\,\mu{\rm G}\;,\;D_0=3\,\times 10^{28}\,{\rm cm^2/s}$, violet line), LOFAR~\cite{Vollmann:2020a} ($B=1\,\mu{\rm G}\;,\;D_0=10^{27}\,{\rm cm^2/s}$, orange line), MWA+GMRT~\cite{Kar:2019} ($B=2\,\mu{\rm G}\;,\;D_0=3\,\times 10^{28}\,{\rm cm^2/s}$, magenta line).  Other relevant campaigns have been conducted with the GBT~\cite{Spekkens:2013,Natarajan:2013,Natarajan:2015} and MWA~\cite{Cook:2020}. Their bounds are not in a suitable form to be shown in \Fig{fig:comp}, but correspond to about $\langle \sigma_a v\rangle \lesssim 10^{-24}{\rm cm^3/s}$ for $M_\chi=100$ GeV and the $b\bar b$ channel.

\subsection{Comparison with $\gamma$-ray analyses}
\begin{figure}[ht!]
\vspace{-2.cm}
   \includegraphics[width=0.49\textwidth]{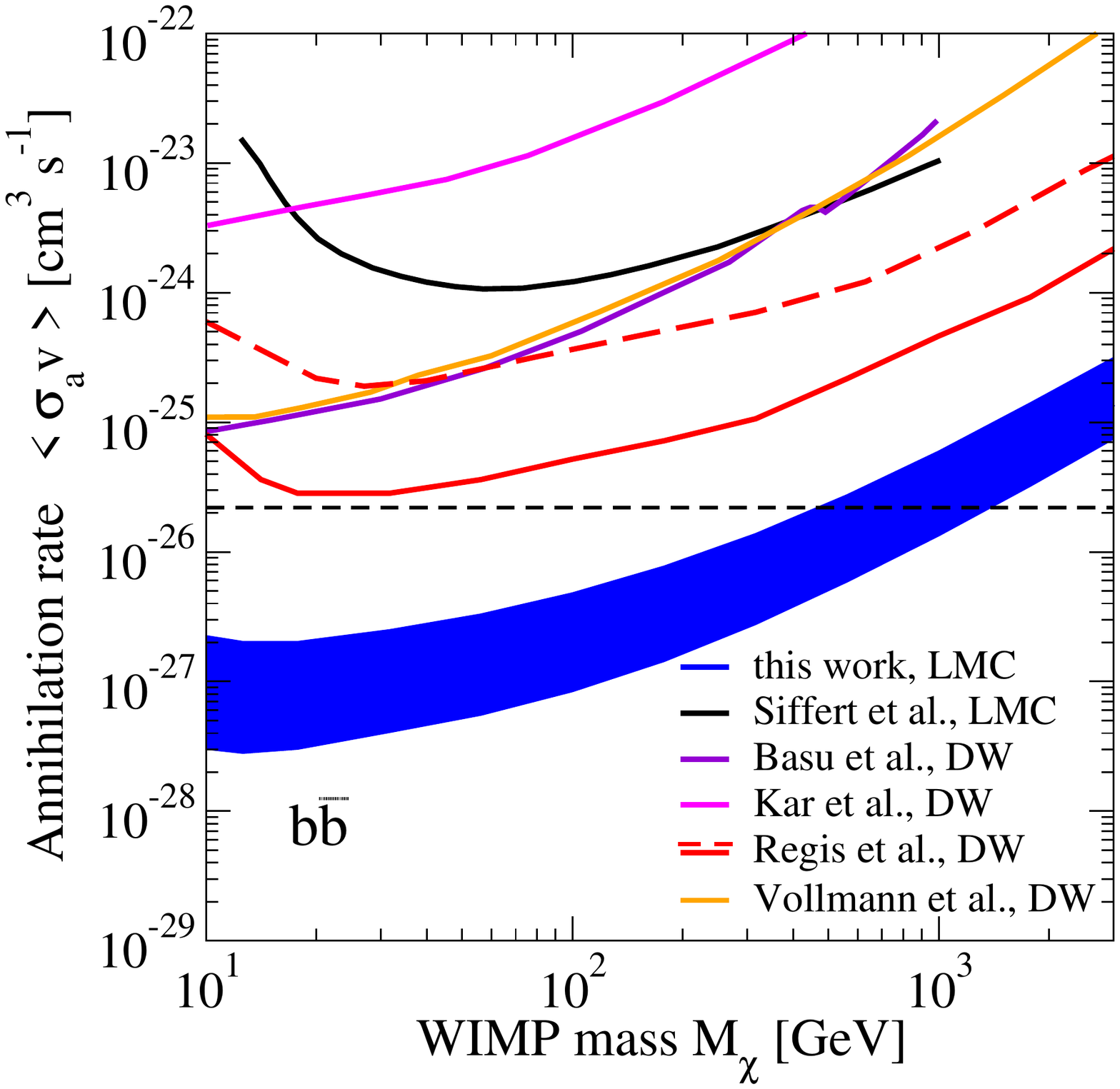}
   \includegraphics[width=0.49\textwidth]{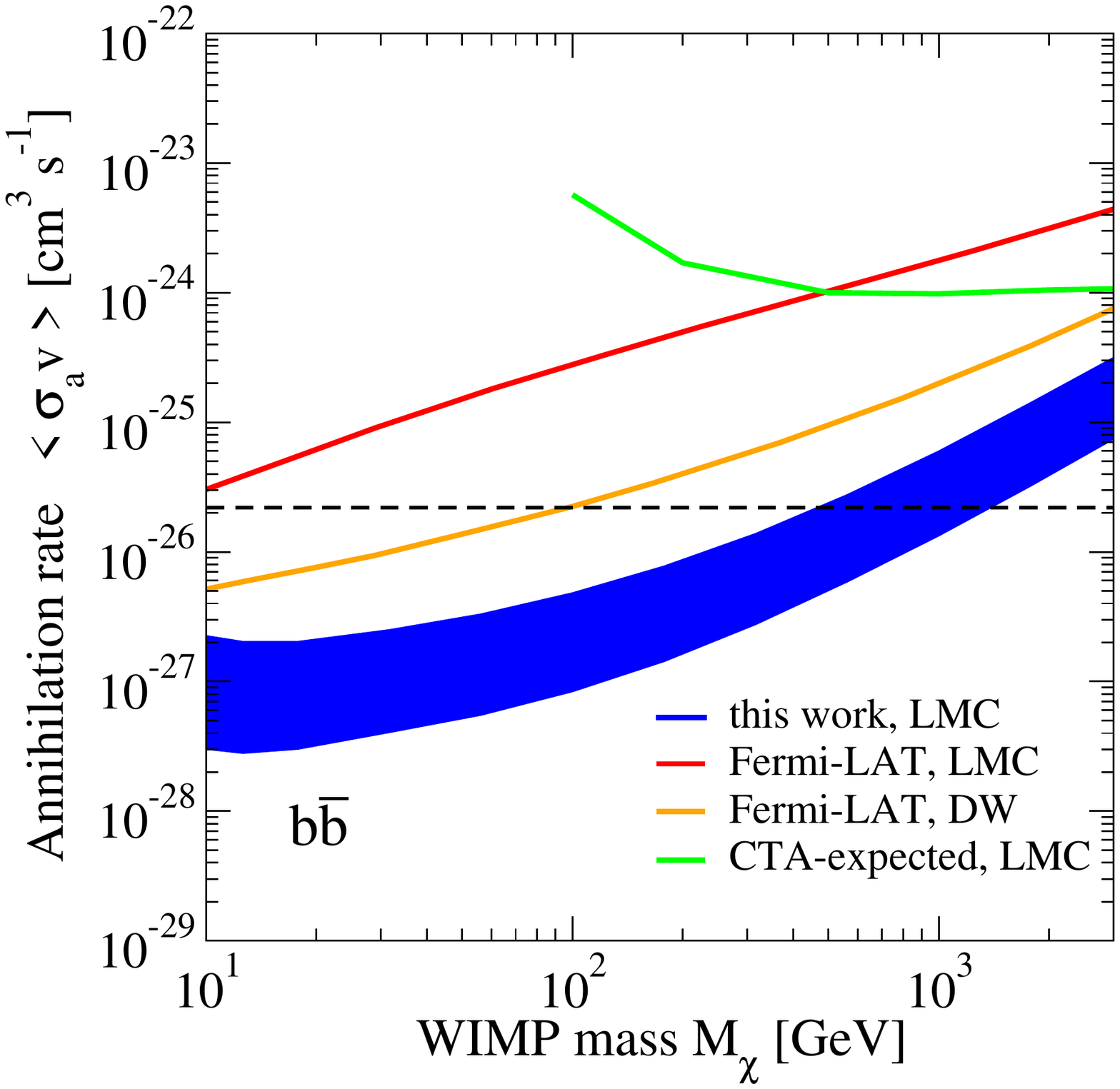}
    \caption{{\bf Left panel:} Comparison with other radio searches, see text for details. ``DW'' labels analyses of dwarf spheroidal galaxies. 
    {\bf Right panel:} Comparison with $\gamma$-ray Fermi-LAT searches on dwarf spheroidals~\cite{Ackermann:2015zua} (orange) and on the LMC~\cite{PhysRevD.91.102001}, and with the expected sensitivity of CTA for LMC~\cite{ 2019EPJWC.20901021I}. 
}
\label{fig:comp}
 \end{figure}

In \Fig{fig:comp} (right), we compare the results of this work with the bounds obtained by the Fermi-LAT Collaboration from the analysis of the LMC~\cite{PhysRevD.91.102001} and dSphs~\cite{Ackermann:2015zua}.
For completeness we also show the expected LMC bounds from the Cherenkov Telescope Array~\cite{ 2019EPJWC.20901021I}, since they can be more constraining than those from Fermi-LAT at high WIMP masses. 

One can immediately see that the LMC radio constraints are much stronger than $\gamma$-ray ones. This should not come as a surprise. Indeed, generically, in WIMP models, the luminosity associated with the induced $\gamma$-rays is comparable to or smaller than that of the injected electrons and positrons. For hadronic channels, the $\gamma$-ray emission mainly proceeds through the production and decay of neutral pions, while electron/positron injection is related to charged pions, and so the two mechanisms are tightly related. For leptonic channels, GeV electrons and positrons have a larger yield than the final state radiation of $\gamma$-ray photons. Therefore, if the cooling time/diffusion length is small enough, so that the energy of the electron/positron is radiated within the source, and the synchrotron loss is the dominant (or, at least, among the most relevant) radiation mechanism in the astrophysical object under investigation, the luminosity produced as synchrotron radiation is comparable to or higher than that from $\gamma$-rays~\cite{Regis:2008ij}. Since radio telescopes are much more sensitive than $\gamma$-ray telescopes for all sources having related mechanisms of emission in the two bands (see, e.g., the level of detail in the ASKAP LMC image compared to the $\gamma$-ray image of the LMC~\cite{TheFermi-LAT:2015lxa}), radio bounds on WIMPs are significantly stronger, {\red when above conditions are satisfied, and} in particular for objects with low astrophysical diffuse radio background, such as the LMC.

The picture is different for dSphs, since there the diffusion length is typically larger than the galaxy itself and the magnetic field is thought to be rather small (and so too the synchrotron radiation), which implies a less favourable ratio between radio and $\gamma$ bounds, even though still comparable (see red solid line in the left panel versus orange line in the right panel).

\Fig{fig:comp} shows that the bound derived in this work is the most stringent bound on WIMPs coming from indirect searches in extragalactic objects.

\section{Conclusions}
\label{sec:conc}
We analysed the ASKAP radio image at 888 MHz of the LMC, in order to search for synchrotron emission induced by WIMP DM annihilations.

The large J-factor of the LMC implies it is one of the best targets for DM indirect searches.
The presence of a magnetic field with strength $>1\,\mu$G makes radio searches in the LMC particularly suited for this purpose.

We detect no evidence for emission arising from WIMP annihilations and derive stringent bounds. Annihilations into leptonic channels provide the most constraining bounds at low masses with the thermal cross-section excluded for masses below 192 GeV ($\tau^+\tau^-$) and 164 GeV ($\mu^+\mu^-$).
Annihilations into quarks and gauge bosons are the most constraining cases at intermediate and high masses with the thermal cross-section excluded below 480 GeV ($b\bar b$) and 358 GeV ($W^+W^-$). 

The comparison with the state-of-the-art in \Fig{fig:comp} shows that the bounds on WIMPs derived in this work are extremely competitive.

We adopted a simple and conservative approach, limiting the analysis to a relatively small region, where simplified assumptions and well-motivated data-driven {\red descriptions can be taken for the
astrophysical ingredients entering the model prediction.
For the two most relevant quantities, the DM spatial profile and the magnetic field, we defined reference models according to observations \cite{Kim:1998,Gaensler:2005}. For the other components (which are important for the computation, but to a somewhat lesser extent), such as the spatial diffusion coefficient, the interstellar radiation fields, and the gas density, we consider their upper or lower limits (all in the direction of minimising the DM signal). 

Our results imply there is little hope to detect a signal from low mass thermal WIMPs in laboratories (i.e., in direct and collider searches), whilst very massive thermal WIMPs remain a viable DM candidate. They can be probed by different techniques, including observations from future radio telescopes, such as the SKA, in particular going to higher frequencies. On a shorter timescale, the addition of short spacing data coming from forthcoming observations}
with the Parkes telescope will provide a complete picture of the LMC at different scales.
With such image at hand, a more refined 3D modeling of the synchrotron emission from the entire LMC can be attempted, with the possibility of further tightening the bounds derived in this analysis.

\section*{Acknowledgement}
The Australian SKA Pathfinder is part of the Australia Telescope National Facility which is managed by CSIRO. Operation of ASKAP is funded by the Australian Government with support from the National Collaborative Research Infrastructure Strategy. ASKAP uses the resources of the Pawsey Supercomputing Centre. Establishment of ASKAP, the Murchison Radio-astronomy Observatory and the Pawsey Supercomputing Centre are initiatives of the Australian Government, with support from the Government of Western Australia and the Science and Industry Endowment Fund. We acknowledge the Wajarri Yamatji people as the traditional owners of the Observatory site.

M.R. would like to thank G. Bernardi and M. Taoso for long-standing discussions on the topic of this work.
M.R. acknowledges support by ``Deciphering the high-energy sky via cross correlation'' funded by the agreement ASI-INAF n. 2017-14-H.0 and by the ``Department of Excellence" grant awarded by the Italian Ministry of Education, University and Research (MIUR).
M.R. and J.R. acknowledge funding from the PRIN research grant ``From  Darklight  to  Dark  Matter: understanding the galaxy/matter connection to measure the Universe'' No. 20179P3PKJ funded by MIUR and from the research grant TAsP (Theoretical Astroparticle Physics) funded by Istituto Nazionale di Fisica Nucleare (INFN). M. B. acknowledges funding by the Deutsche Forschungsgemeinschaft (DFG, German Research Foundation) under Germany’s Excellence Strategy – EXC 2121 ‘Quantum Universe’ – 390833306.

\bibliographystyle{hunsrt}
\bibliography{biblio}

\appendix
\section{Consistency checks}
\label{sec:appA}
In this Appendix, we describe a few consistency checks we performed.

\begin{figure}[t]
\vspace{-2.cm}
\centering
   \includegraphics[width=0.75\textwidth]{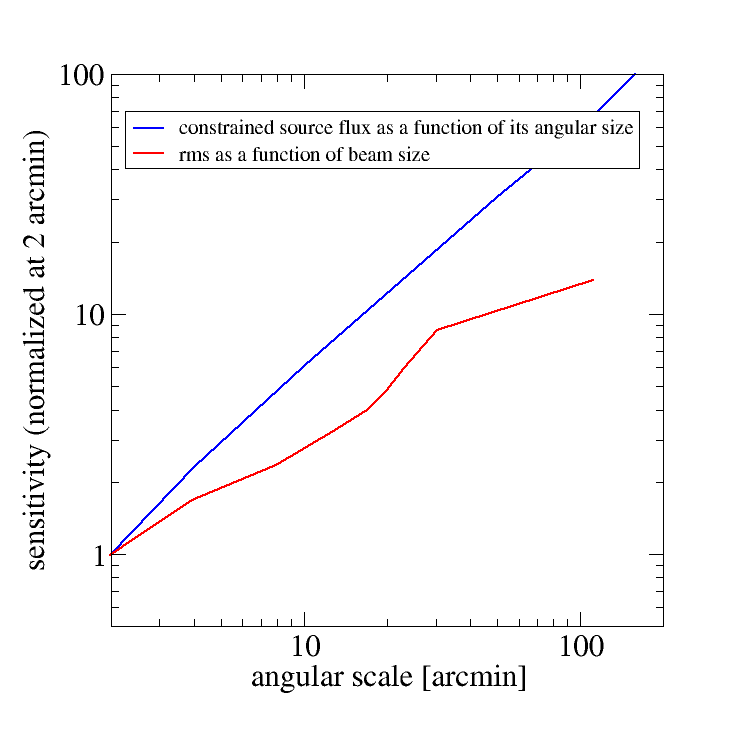}
    \caption{The red line shows the rms sensitivity of the LMC image as a function of the tapering size. The blue curve shows the total flux excluded by the analysis described in the main text as a function of the size of the source. Both curves are normalized to one at two arcmin.}
\label{fig:sensitivity}
 \end{figure}

First, a key and not obvious (for an interferometric image) point is the actual sensitivity of the image to large scale diffuse emissions.
In order to understand how the sensitivity scales as a function of the size of the source, we taper the visibilities by different angular scales, generate an image with robust weighting, and then measure the standard deviation in the image.
For technical reasons and since here we are mainly interested in understanding the trend but not the absolute value, we performed the analysis on Stokes V and considering one of the LMC pointings (all pointings were taken at similar directions and times).
The rms sensitivity normalized to one at 2 arcmin is plotted in \Fig{fig:sensitivity} (red line) as a function of the tapering size.

In the same figure, the blue line reports the total flux of a source which is excluded at 95\% C.L. by the analysis described in the main text, as a function of the angular scale of the source. Again values are normalized to one for a source with FWHM$=2$ arcmin. One can quickly check that, in the case of uniform rms, no masking and no confusion, \Eq{eq:like} would imply a linear scaling. The blue curve is derived by considering Gaussian sources of different widths (essentially replacing the DM component with a Gaussian emission and then repeating all the steps described in Sects.~\ref{sec:model} and \ref{sec:res}). The actual behaviour is close to linear scaling.

The bottom-line of \Fig{fig:sensitivity} is that the ``theoretical'' degradation of the sensitivity as the angular scale of the sources increases, as assumed by the analysis described in the paper, is higher if compared to the results obtained through the tapering test. This ensures our bounds are conservative.

\begin{figure}[t]
\vspace{-2.cm}
   \includegraphics[width=0.49\textwidth]{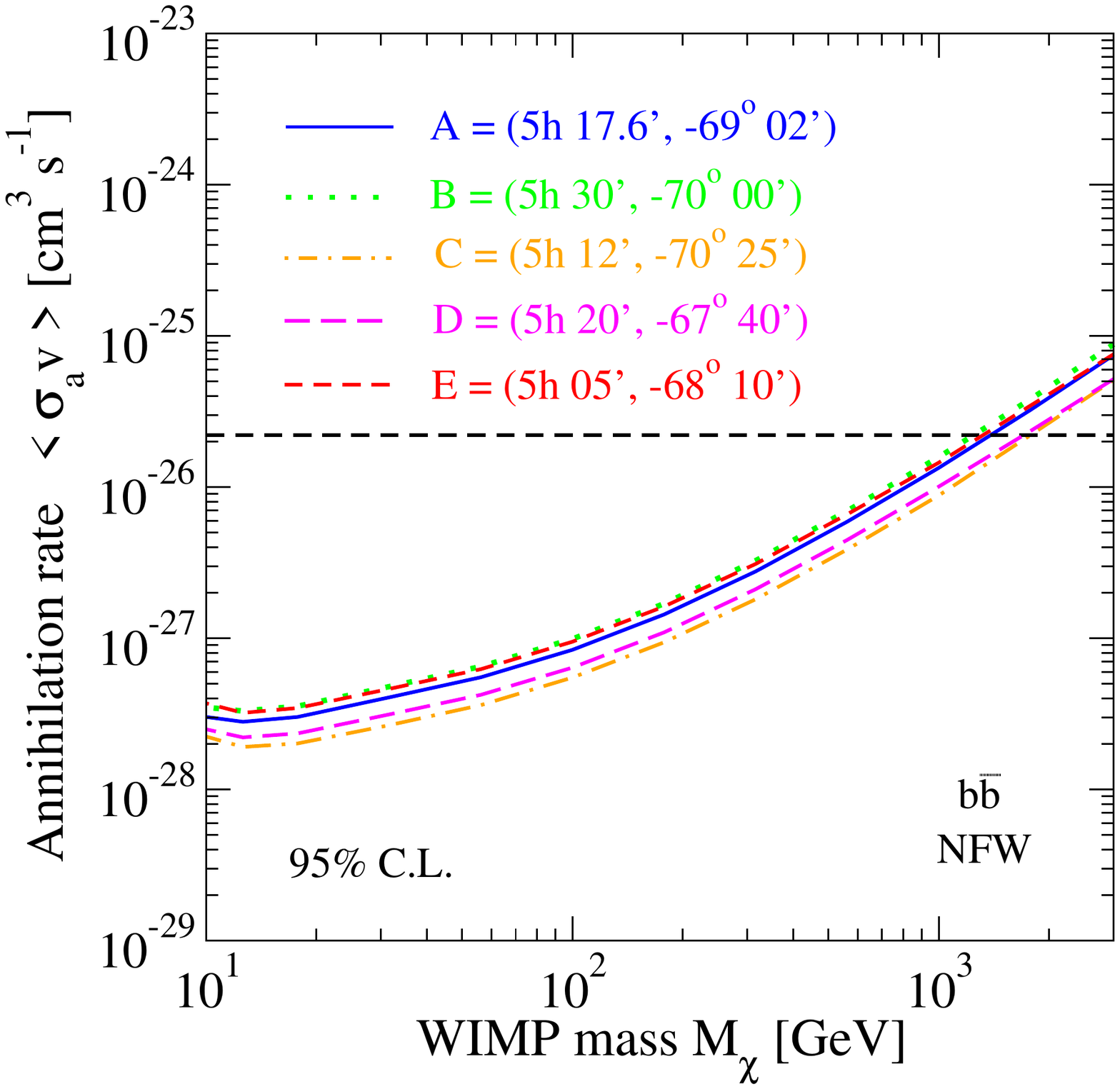}
   \includegraphics[width=0.49\textwidth]{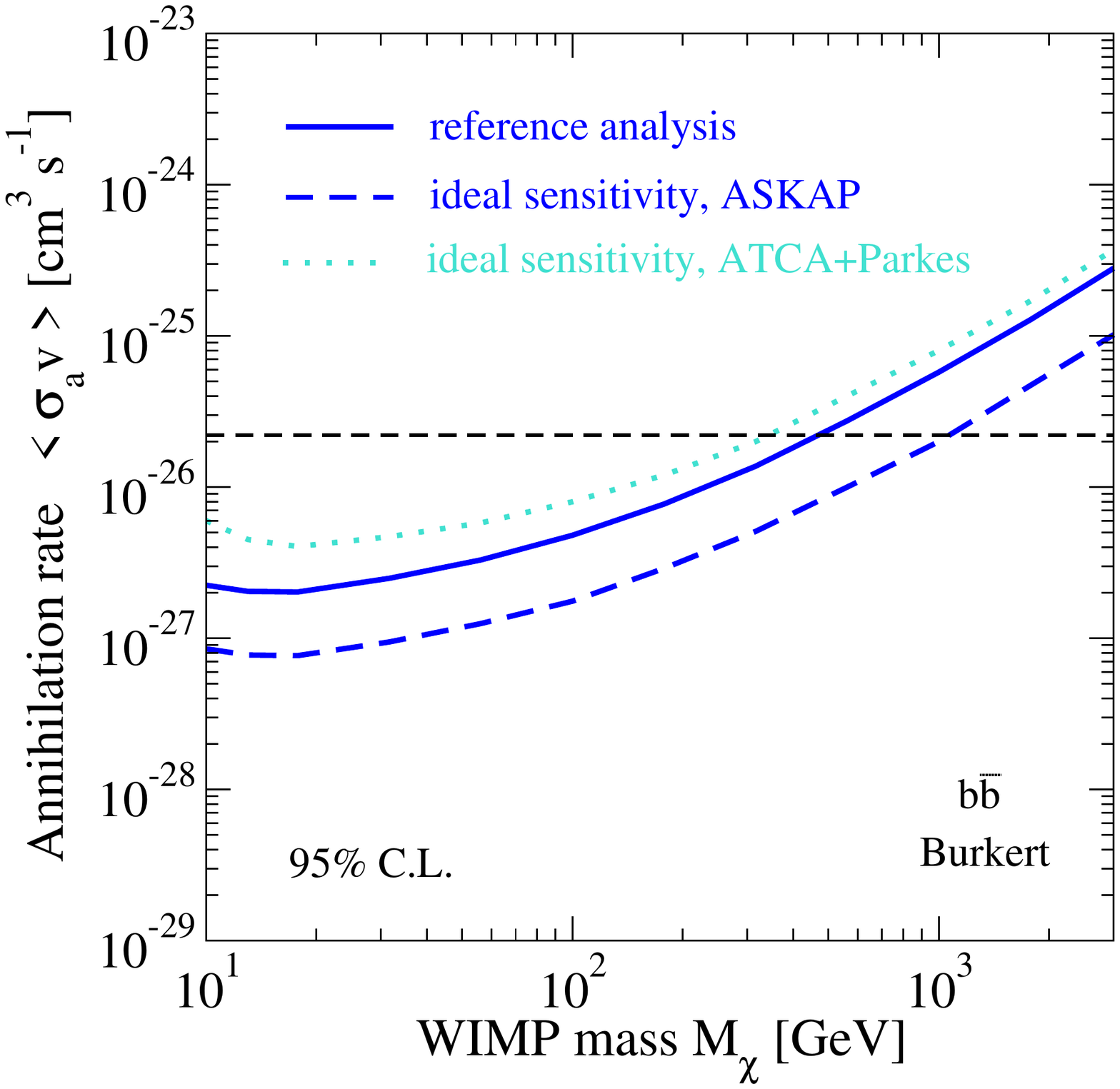}
    \caption{{\bf Left panel:} Comparison of the bounds obtained by centering the DM distribution on different positions across the map {\red (labeled as A, B, C, D and E to easy their reference in the text)}.
    {\bf Right panel:} Comparison of the bounds obtained in the reference analysis of this work (solid) with the bounds one would obtain considering the maximal ideal sensitivity of the EMU image (dashed) and analysing the ATCA+Parkes image of Ref.~\cite{Hughes:2007} (dotted).
}
\label{fig:checks}
 \end{figure}

To test against possible systematics associated to the selected region of the image, we re-do the same analysis outlined in the main text, but now centering the DM distribution on different positions across the map. Since we are dealing with a non-detection, they should all provide similar bounds, because the RMS sensitivity is approximately uniform across the map.
We compare five different positions, as listed in \Fig{fig:checks} (left).
Concerning the DM model we take the same description as in Section~\ref{sec:model}, but centered in the new positions. This is clearly not realistic, but functional to our test.
We find that positions that are far from the LMC disk (i.e., $C$ and $D$ in \Fig{fig:checks}) provide slightly more constraining bounds, with the component $S_{disc}$ compatible with zero. Positions located on the LMC disk (i.e., $B$ and $E$ in \Fig{fig:checks}) lead to bounds similar to the ones described in the main text (case $A$), with the fit requiring a disk component different from zero.
\Fig{fig:checks} (left) reports the bound in the case of the NFW profile and annihilation into $b\bar b$.

In \Fig{fig:checks} (right) we compare our results with the maximal sensitivity that can be achieved with the image we have at hand. The latter is derived by keeping the original resolution (FWHM=13'') and evaluating $\chi^2=\sum_{i=1}^{N_{pix}} \left(\frac{S_{DM}^i}{\sigma_{rms}^i}\right)^2/N_{pix}^{FWHM}$, which can be seen as setting to zero all pixels in the map (after RMS determination).
We show the result for the Burkert profile since it is the most extended case, so where the number of pixels relevant for the $\chi^2$ determination is largest, which implies the sensitivity difference is largest.
As expected the bound derived with such ideal sensitivity is more constraining than for our reference analysis, but by a rather limited factor (between 2 and 3).

In the same Figure, we derive WIMP bounds from the LMC map at 1.4 GHz presented in Ref.~\cite{Hughes:2007}.
Such map contains all scales above $40''$ being a combination of ATCA and (single-dish) Parkes data, contrary to the ASKAP image having only interferometric data. This means that the very large-scale emission would need a more careful treatment than the simple model introduced in Section~\ref{sec:model}. For this reason, we perform the comparison in the ``ideal-sensitivity'' case.
The analysis of ATCA+Parkes data is performed in the same way as for the ASKAP map.
From the ratio of the RMS sensitivity of the two maps (300 versus 58 $\mu$Jy/beam), and considering the different frequency (1.4 versus 0.888 GHz) and beam (40'' versus 13''), we expect the ATCA+Parkes bound to be a factor around 5 weaker than the ASKAP one.
Results are along the line of expectations, see dotted versus dashed lines, providing a consistency check.

\end{document}